\documentclass[a4paper,11pt]{article}
\usepackage{jcappub}

\usepackage{array}
\usepackage{mathtools}

\usepackage{etoolbox}

\title{The Quantum Origin of Quasi de Sitter: a Model Independent Quantum Cosmological Tilt}
 
 \author[a]{C\'esar G\'omez,} 
\emailAdd{cesar.gomez@uam.es}

\author[b,c]{Raul Jimenez} 
\emailAdd{raul.jimenez@icc.ub.edu}

\affiliation[a]{Instituto de F\'{i}sica Te\'orica UAM-CSIC, Universidad Aut\'onoma de Madrid, Cantoblanco, 28049 Madrid, Spain.}
\affiliation[b]{ICC, University of Barcelona, Marti i Franques 1, 08028 Barcelona, Spain.}
\affiliation[c]{ICREA, Pg. Lluis Companys 23, Barcelona, E-08010, Spain.}

\abstract{
The most robust prediction of inflationary cosmology is the existence of a red tilt for the spectrum of curvature fluctuations that is experimentally of order $0.04$. The tilt is derived solving the exact equation for quantum fluctuations in a quasi de Sitter background defined by a equation of state $\epsilon \equiv \frac{(p+\rho)}{\rho}$ with $\epsilon$ small but non vanishing. The experimental data selects among the different quasi de Sitter inflaton potentials. The origin of the lack of scale invariance associated with the tilt is however classical in essence and parametrized by the slow roll of the inflaton potential. Here we present a purely quantum mechanical and model independent derivation of the tilt. This derivation is based on two basic observations: first, the correlator for gauge invariant variables is related to the {\it quantum Fisher function} measuring the quantum dependence of the family of pure de Sitter vacua on the energy scale parameter; second, this quantum Fisher function has a non vanishing scale dependent red tilt that, at the energy scales of physical interest, fits the effective quasi de Sitter prediction as well as the experimental value. This is a result that is model independent and only based on the quantum features of the family of de Sitter vacua.  
}

\begin{document} 
\maketitle

\section{Introduction}
One of the most fascinating results of modern Cosmology is that of a quantum origin of the large scale structure of the Universe \cite{Mukhanov,Hawking,Guth:1982ec,Starobinsky:1982ee,Bardeen:1983qw}. This involves two basic ingredients: an expanding quasi de Sitter background geometry \cite{Starobinsky,Sato:1980yn,Guth,Linde:1981mu,Albrecht,Linde:1983gd} with equation of state $\epsilon \equiv \frac{(p+\rho)}{\rho} << 1$, where $p$ is the pressure and $\rho$ the density in the corresponding energy-momentun tensor,  but non-vanishing, and scalar quantum field theory in this curved background. The quasi de Sitter background defines a {\it classical model} of {\it ``cosmological constant decay"} in physical time, something needed to end inflation. In this quasi de Sitter background, quantum fluctuations of the scalar field induce curvature fluctuations that {\it after horizon exit  classicalization} and subsequent {\it gravitational amplification} can seed the large scale structure we measure in the cosmic microwave background (CMB) and large scale structure. This beautiful framework, that combines a classical model of what we can superficially describe as cosmological constant decay with standard quantum field theory in an expanding curved space-time, leads to a set of robust experimental predictions. In particular, fluctuations are nearly gaussian and a departure from exact scale invariance measured by the tilt $n_s-1= -3\epsilon + \frac{d\ln \epsilon}{dN}$ where the second term measures the time dependence of epsilon in units of e-foldings $N$. 
The tilt that measures the departure from exact scale invariance is the direct consequence of the classical decay of the early inflaton potential energy. This decay is modeled by the corresponding inflaton potential. The solutions to the equation for the quantum fluctuations in this decaying background carry the tilt dependence. The quantum fluctuations are defined relative to a quantum vacuum state that is generically
 a quantum squeezed state supporting a lot of quantum entanglement~\cite{Grishchuk:1990bj,Grishchuk:1992tw}.

There are two obvious problems we can pose to the former scheme. The most important one is the intrinsically classical way to describe the "decay of the primordial dark energy". This leads to a plethora of potential different models of inflation with a set of associated methodological problems going from fine tuning issues to the potential existence of a multiverse~\cite{Loeb}. The other problem is how much quantum the whole approach is (see e.g. Ref.~\cite{Starobinsky:1986fx,Polarski:1995jg,Lesgourgues:1996jc,Egusquiza:1997ez,Kiefer:1998qe}). More specifically, if we can extract from CMB measurements the intrinsic quantum features of the state used to define the different correlators defining the CMB temperature fluctuations. If the quantum features are too small and exponentially suppressed by the duration of inflation\cite{dePutter}, we can always wonder if we cannot model with a classical probability distribution all the present observations~\cite{Berera:1995ie,Berera:1995wh,Steinhardt,Brandenberger,LopezNacir:2011kk,Senatore:2011sp}. This line of reasoning leads to discuss cosmological Bell inequalities~\cite{Campo:2005sv,Maldacena:2015bha} that, unfortunately, for most realistic models are too small to be detected as well as quantum discord and entanglement entropy (see Ref.~\cite{Martin:2016tbd,Brahma:2020zpk} for recent discussions).
In this article we will concentrate our attention on the first issue, namely the classical description of the primordial dark energy decay. 

The observational sign of the decay of the primordial dark energy, what we normally describe with a quasi de Sitter phase characterized by small slow roll parameters, is the {\it red tilt} of the power spectrum of curvature fluctuations. What we will present in this paper is {\it a purely quantum mechanical and model independent approach} to extract this tilt. In a nutshell the idea is as follows. For pure de Sitter characterized by a given value of the Hubble parameter $H$, we define the different de Sitter vacua, normally known in the literature as $\alpha$-vacua (see Ref.~ \cite{Chernikov:1968zm,Bunch:1978yq,Allen:1985ux,Mottola:1984ar,Bousso:2001mw,Anderson:2017hts,Danielsson:2018qpa,Danielsson:2002kx,Danielsson:2002mb,Danielsson:2004xw,Polyakov:2007mm,Polyakov:2009nq,Krotov:2010ma,Polyakov:2012uc,Anderson:2013ila,Anderson:2013zia} for definitions and different discussions on vacuum selection in de Sitter). These set of vacua can be parametrized by an energy scale $|\Lambda\rangle$. Once we count with the quantum mechanical representation of these states, we can define as a measure of the dependence on the energy scale the corresponding {\it quantum Fisher information} that we will denote ${\cal{F}}_Q(\Lambda)$ (For definitions and general theory of quantum information see Ref.~\cite{Paris} and for recent applications to cosmology see Ref.~\cite{GJ1,GJ2,GJ3}). What this information is measuring is the square uncertainty of the generator defining $|\Lambda\rangle$ vacuum transformations. More precisely, the quantum Fisher measures the $\Lambda$ dependence of the quantum phases defining the state $|\Lambda\rangle$. Note that this quantum Fisher only depends on the quantum features of de Sitter and therefore is completely model independent. The generator of $|\Lambda\rangle$ vacuum transformations can be expressed in terms of gauge invariant Mukhanov-Sasaki~\cite{Mukhanov,Kodama:1985bj} variables and the corresponding quantum Fisher depends on $\Lambda$ with a uniquely defined {\it tilt} that measures {\it quantum departures} from scale invariance. 

Once we set up the conceptual frame the key result of this article can be summarized as follows:

{\it The quantum Fisher tilt at scale $\Lambda$ defines the tilt of the power spectrum of curvature fluctuations at the corresponding quasi de Sitter scale.}

\begin{figure}
\centering
\includegraphics[width=.9\columnwidth]{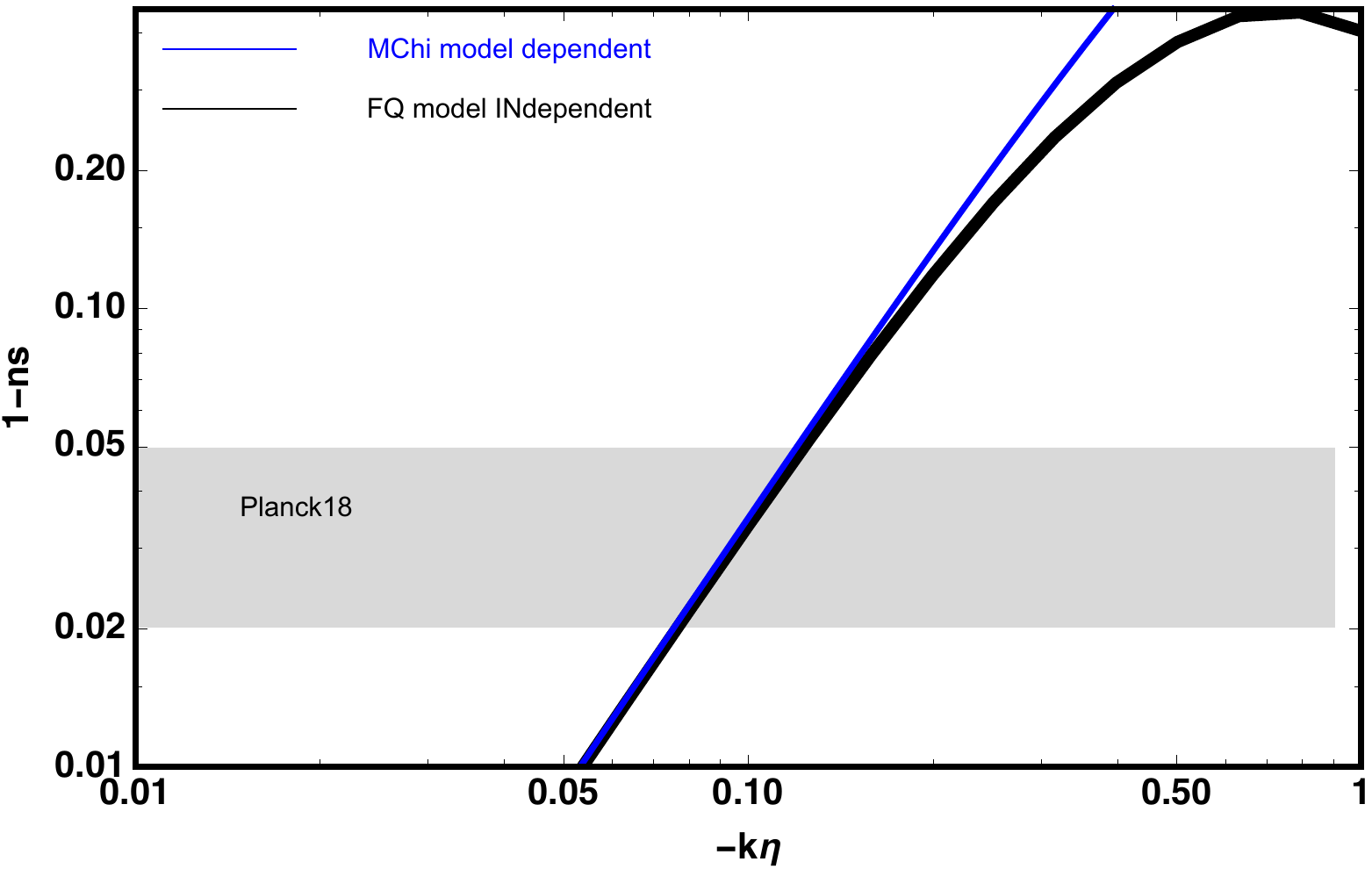} 
\caption{The black line shows the tilt in the power spectrum of primordial fluctuations as a function of scale computed from the quantum Fisher ${\cal{F}}_Q$. Note that for scales a factor ten smaller than the current horizon, $1-n_s \sim 0.04$. Overplotted as a blue line is the predicted value from Mukhanov-Chibisov~\cite{Mukhanov}. Note that the slope is the same as our derived value. The original tilt in MChi can be related to the quantum Fisher tilt by simply identifying MCH scale M with our scale $\Lambda$. More precisely ${\rm MChi}_{\rm tilt(M)} = {\cal FQ} _{\rm tilt (k \eta = 4 M/H)}$. The gray rectangle depicts the current allowed values for $1-n_s$ at 99\% confidence from Planck18~\cite{Planck18}.}
\label{fig:ns}
\end{figure}

To make this statement more transparent, we compare the original tilt discovered by Mukhanov and Chibisov~\cite{Mukhanov} depending on the {\it quasi de Sitter scale} $M$ (following their original notation) and the quantum model independent tilt defined by the quantum Fisher information as a function of the energy scale $\Lambda$ defining the vacuum. The agreement of both pictures in energy scales around $H$ is presented in Fig.~\ref{fig:ns}. 

In summary, we conclude that the quantum Fisher information of de Sitter $|\Lambda\rangle$ vacua contains the full information about the red tilt of quantum fluctuations responsible for large scale formation. Moreover, this tilt can be extracted without making any model dependent assumption on the classical features of a quasi de Sitter phase. In essence, this quasi de Sitter phase is simply (for what concerns the tilt) an effective description of the energy dependence of the quantum phases of the $|\Lambda\rangle$ de Sitter vacuum states. 

Before going into the detailed derivation of the quantum tilt let us make a general comment on the connection of the results presented in this note and the most general issue of de Sitter quantum stability. As already pointed out, a quasi de Sitter phase can be interpreted as an effective description of de Sitter instability. The microscopic approach to the quantum instability of de Sitter based on a coherent state picture with a quantum breaking time was developed in \cite{Dvali:2013eja,Dvali:2014gua,Dvali:2017eba}. Here we are only interested in fishing out a very concrete and experimentally observable sign of this potential instability, namely the quantum departure from scale invariance as defined by the tilt. Our message is that the quantum Fisher information for the $|\Lambda\rangle$ vacua can be interpreted as the fine grained information about the quantum effects underlying the existence of a non vanishing red tilt.

\section{Primordial power spectrum}
The key notion to define the primordial power spectrum of curvature fluctuations is the gauge invariant Mukhanov-Sasaki variable. For the Fourier modes $v_{k}$, for $k$ the comoving momentum and at conformal time $\eta$, the correlator $\langle v_{k}v^*_{-k}\rangle$ is given by\footnote{In what follows we will ignore the numerical factors that don't affect the value of the tilt.}
\begin{equation}
\langle v_{k}v^*_{-k}\rangle = \frac{1}{k^3}{\cal {P}}_{v}(k)
\end{equation}
with
\begin{equation}
{\cal {P}}_{v}(k) = k^3 |f_k|^2
\end{equation}
with $f_k(\eta)$ the solution of
\begin{equation}\label{equation}
f_k^{''} +(k^2-\frac{z^{''}}{z})f_k=0
\end{equation}
with $z=a \sqrt{\epsilon}$ and $\epsilon$ defined by the state equation
\begin{equation}
\epsilon = \frac{(p+\rho)}{\rho}
\end{equation}
In this sense we get
\begin{equation}
\langle v_{k}v^*_{-k}\rangle(\eta;\epsilon) = |f_k(\eta;\epsilon)|^2
\end{equation}
for $f_k(\eta;\epsilon)$ the solution to equation (\ref{equation}). This equation can be easily solved for pure de Sitter i.e. $\epsilon=0$ leading for super horizon modes $-k\eta << 1$ to
\begin{equation}
|f_k|^2(\eta) \sim \frac{\eta}{(k\eta)^3}
\end{equation}
that yields
\begin{equation}
k^3\langle v_{k}v^*_{-k}\rangle(\eta;\epsilon=0) = {\cal {P}}_{v}(k,\eta)= \frac{1}{\eta^2}
\end{equation}
which is exactly scale invariant. To account for the effect of $\epsilon$ we need to solve (\ref{equation}) for $\epsilon\neq 0$. This can be done using the matching with Bunch-Davis modes in the region $-(k\eta)>>1$. The result, for superhorizon modes is given by
\begin{equation}
k^3\langle v_{k}v^*_{-k}\rangle(\eta;\epsilon) \sim \frac{1}{\eta^2}\frac{1}{(k\eta)^{\alpha(\epsilon)}}
\end{equation}
leading to the tilt
\begin{equation}
1-n_s= \alpha(\epsilon)
\end{equation}
Parametrizing $\frac{z^{''}}{z} = \frac{\beta(\beta+1)}{\eta^2}$ this leads to $1-n_s = \alpha =-( 2\beta+4)$ with $\beta =-2$ for pure de Sitter. Note that in our convention $\alpha$ is positive for {\it red tilt}.
This is the famous departure from scale invariance induced by a non vanishing value of $\epsilon$. Here is important to note that this anomalous scale invariance is due to the {\it classical} quasi de Sitter design of the cosmological constant decay as defined by the slow roll parameter $\epsilon$ while the exact form of the anomalous scale invariance is dictated by the equation (\ref{equation}) governing the quantum fluctuations in a quasi de Sitter classical background \footnote{The Mukhanov Chibisov tilt depicted in Fig.~\ref{fig:ns} corresponds to the particular model $\beta = \frac{3}{2}\sqrt{1+\frac{4}{9} \frac{(4M)^2}{H^2}}$. The tilt is implemented in the order of the Besell function that is $\beta+1/2$.} 

Usually, a new variable $\chi_k$ defining the curvature perturbations is introduced as\footnote{This variable is defined in terms of the Bardeen potential $\Phi$ as $\chi= \frac{2({\cal{H}}^{-1} \Phi' +\Phi)}{3(1+\omega)} + \Phi$ with $p=\omega \rho$ and ${\cal{H}} = \frac{a'}{a}$.}
\begin{equation}
v_k= a M_P \sqrt{\epsilon} \chi_k
\end{equation}
leading to
\begin{equation}
\langle \chi_{k}\chi_{-k}\rangle = \frac{1}{a^2\epsilon M_P^2} \langle v_{k}v^*_{-k}\rangle(\eta;\epsilon)
\end{equation}
or equivalently to
\begin{equation}
k^3 \langle \chi_{k}\chi_{-k}\rangle =\frac{1}{a^2\epsilon M_P^2}\frac{1}{\eta^2}\frac{1}{(k\eta)^{\alpha(\epsilon)}} \sim \frac{H^2(k)}{M_P^2\epsilon(k)}\frac{1}{(k\eta)^{\alpha(\epsilon)}}
\end{equation}
which is the well known tilt dependence of the curvature power spectrum. It is customary to write the curvature power spectrum around a {\it pivot} scale $k_0$ as
\begin{equation}\label{power}
\frac{H^2(k_0)}{M_P^2\epsilon(k_0)}(\frac{k}{k_0})^{\alpha(\epsilon)}
\end{equation}
corresponding to use $k\eta = \frac{k_0}{k}$ i.e. $\eta =\frac{1}{k}$ with $k$ near the pivot scale $k_0$. In other words (\ref{power}) gives us the power spectrum around the horizon exit point $k\eta =1$.

For convenience, let us introduce a function ${\cal{F}}$ defined as
\begin{equation}
{\cal{F}} = \frac{\eta^2}{(k\eta)^6} \frac{1}{(k\eta)^{2\alpha}}
\end{equation}
that leads to
\begin{equation}
 \langle \chi_{k}\chi_{-k}\rangle = \frac{1}{M_P^2 a^2 \epsilon} \sqrt{{\cal{F}}}
 \end{equation}
 i.e.
 \begin{equation}
 \langle v_{k}v_{-k}\rangle(\eta;\epsilon) = \sqrt{{\cal{F}}}
 \end{equation}
The rest of the paper will be dedicated to unveil the {\it quantum information meaning} of the function ${\cal{F}}$ as:

 {\it The  quantum Fisher function measuring the $\Lambda$ dependence of pure de Sitter $|\Lambda\rangle$ vacua.} 
 
 In order to do that we will review in the next section the notion of $|\Lambda\rangle$ vacuum for pure de Sitter.

\section{de Sitter vacuum}
Let us start defining the Bunch-Davis algebra of creation annihilation operators $a_k^{BD}, a_{-k}^{\dagger BD}$ for $k$ the comoving momentum. The BD vacuum, normally known as the euclidean vacuum, is defined by
\begin{equation}
a_k^{BD}|E\rangle =0
\end{equation}
for all values of $k$. 

The family of $|\alpha\rangle$ vacua, for $\alpha$ a generic complex number is defined by a family of Bogolioubov transformations
\begin{equation}
a_k(\alpha) = A(\alpha) a_k^{BD} +B(\alpha) a_{-k}^{\dagger BD}
\end{equation}
satisfying
\begin{equation}
|A(\alpha)|^2 -|B(\alpha)|^2 =1
\end{equation}
The $|\alpha\rangle$ vacuum is defined by the condition
\begin{equation}
a_k(\alpha)|\alpha\rangle =0
\end{equation}
for all values of $k$. Let us define an {\it energy scale } $\Lambda$ and the Bogolioubov transformation
\begin{equation}
A(\Lambda) = \cosh( r(\Lambda))
\end{equation}
and 
\begin{equation}
B(\Lambda) = - e^{2i\phi(\Lambda)} \sinh r(\Lambda)
\end{equation}
with $r(\Lambda) = - \sinh^{-1}(\frac{H}{2\Lambda})$ and $\phi(\Lambda) = -\frac{\pi}{4} -\frac{1}{2} \tan^{-1}(\frac{H}{2\Lambda})$. 

Defining the complex number $\alpha$  by
 \begin{equation}
 \alpha= \ln \tanh(r(\Lambda)) -2i \phi(\Lambda)
 \end{equation}
 the Bogoloubov transformation defined above becomes
 \begin{equation}
 a_k(\alpha) = N(\alpha)(a_k^{BD} - e^{\alpha^*}a_{k}^{\dagger BD}) \end{equation}
 where $N(\alpha) = \frac{1}{\sqrt{1-e^{\alpha+\alpha^*}}}$
 and the state $|\alpha\rangle$ is given by
 \begin{equation}
 |\alpha\rangle = Ce^{\sum_k\frac{1}{2}e^{\alpha^*}a_k^{\dagger BD}a_{-k}^{\dagger BD}}
\end{equation}
with $C$ a normalization factor.
Since the energy scale $\Lambda$ fixes the value of $\alpha$ we can associate the vacuum $|\alpha\rangle$ with the corresponding energy scale $\Lambda$. Thus from now on we will use the energy scale $\Lambda$ to refer to the $|\Lambda\rangle$ vacuum.

For a given $\Lambda$ we can define for a given value of $k$ the state $|k,\eta\rangle$ as
\begin{equation}
|k, \eta\rangle = e^{\frac{1}{2}e^{\alpha^*}a_k^{\dagger BD}a_{-k}^{\dagger BD}}|E\rangle
\end{equation}
with $\eta= -\frac{\Lambda}{kH}$. In this notation we get
\begin{equation}\label{state}
|\Lambda\rangle = \prod_k |k, \eta= -\frac{\Lambda}{kH}\rangle
\end{equation}

Some remarks are relevant:
\begin{itemize}
\item For $\Lambda=\infty$ the vacuum $|\Lambda=\infty\rangle$ is the BD vacuum. This explains the so called trans-Planckian puzzle. Indeed the BD vacuum is defined at infinite energy scale i.e. much larger than the Planck mass $M_P$. 
\item The state $|\Lambda\rangle$ for $\Lambda$ finite depends on different values of the conformal time i.e. cannot be associated with a given value of time.
\item The special value $\Lambda=0$ is concentrated on $\eta=0$ and corresponds to $\alpha = -i\pi$. Another interesting state is $\Lambda =H$ where all the comoving momentum $k$ in (\ref{state}) are defined at horizon exit i.e. $k\eta =1$.
\item Note that these states, with the exception of the BD vacuum $\Lambda = \infty$, are not invariant under {\it scale transformations}. 
\end{itemize}

\subsection{A macroscopic view}
Let us consider quantum states of the type
\begin{equation}
|n(k),n(-k)\rangle
\end{equation}
with the same distribution for modes $k$ and $-k$. We can define a macroscopic basis $|N\rangle$ with $N= \int dk n(k)$. From the construction of the state $|\alpha(\Lambda)\rangle$ it is clear that this state can be represented as
\begin{equation}\label{repre}
|\alpha(\Lambda\rangle = C \sum_{N} \tanh (r(\Lambda))^N e^{iN\phi(\Lambda)} |N\rangle
\end{equation}
with $C$ a normalization factor. In the next section we will use this representation to evaluate the quantumness of these states.

 \section{de Sitter vacuum and quantum information}
 Let us consider the de Sitter vacuum $|\Lambda\rangle$ as given in the macroscopic basis (\ref{repre}). Let us now compute the quantum Fisher information relative to the scale parameter $\Lambda$. This quantum Fisher is defined as
 \begin{equation}
 F_Q(\Lambda) = \sum_N N^2 (\frac{\partial \phi(\Lambda)}{\partial (k\eta)})^2 \tanh(\Lambda)^{2N} - (\sum_N N (\frac{\partial \phi(\Lambda)}{\partial (k\eta)}) \tanh(\Lambda)^{2N})^2
 \end{equation}
 This can be easily evaluated leading to
 \begin{equation}
 F_Q(\Lambda) = \frac{1}{(k\eta)^{8-2\alpha(k\eta)}}
 \end{equation}
 for $\Lambda =k\eta H$.
This is the quantum Fisher relative to variations of $k\eta$ i.e. measuring the dependence of the quantum phase on $k\eta = \frac{\Lambda}{H}$. 
 The quantum Fisher relative to the new variable $ \frac{1}{\eta}$ differs by a factor $k^2 \eta^4$ and is given by
  \begin{equation}\label{Fisher}
 {\cal{F}}_Q(\eta, \Lambda) = \frac{k^2 \eta^4}{(k\eta)^{8-2\alpha(k\eta)}}
 \end{equation}
The next task is to relate ${\cal{F}}_Q(\eta, \Lambda)$ with the correlator for the Mukhanov-Sasaki gauge invariant variables  $\langle v_{k}v^*_{-k}\rangle(\eta;\epsilon)$. At this point the reader can wonder how the quantum Fisher computed in pure de Sitter can have any information on the parameter $\epsilon$ that defines a classical departure from de Sitter. To answer this question let us proceed as follows.

First of all, the quantum Fisher is formally defined for the operator generating transformations of the $\Lambda$ vacua. Once written in terms of BD creation annihilation operators this generator can be transformed into $\langle v_k v^*_{-k}\rangle$. The key difference with respect to the generator of transformations in conformal time $\eta$ is to replace $\eta$ by $\frac{z'}{z}= \frac{1}{\eta}$. Recall that it is this quantity the one defining the "coupling" of the interaction Hamiltonian. Thus we get the following correspondence:
\begin{center}
$\langle v_k v^*_{-k}\rangle$ $\Rightarrow$ $\sqrt{{\cal{F}}_Q}$ 
\end{center}
The reason for the square root is that the quantum Fisher refers to the square of the generator. Using (\ref{Fisher}) this correspondence leads to the standard expression for $\langle v_k v^*_{-k}\rangle$ namely
\begin{equation}
\langle v_k v^*_{-k}\rangle (\eta, \epsilon) = \sqrt{{\cal{F}}_Q(\eta,\Lambda)}
\end{equation}
provided the {\it physical tilt $\alpha(\epsilon)$ is given by the quantum Fisher tilt $\alpha(\Lambda)$} i.e.
\begin{equation}
1-n_s = \alpha
\end{equation}
Thus we get the important result:

{\it The physical tilt of the primordial power spectrum that normally depends on the classical slow roll input $\epsilon$ is encoded in the quantum tilt of the quantum Fisher function for the $\Lambda$ vacuum of pure de Sitter through the correspondence:
\begin{equation}\label{correspondence}
\alpha_{qdS}(\epsilon) \Rightarrow \alpha_{Fisher}(\Lambda)
\end{equation}
}

Note that this relation associates the quasi de Sitter scale at which we define the phenomenological parameter $\epsilon$ with the vacuum de Sitter energy scale $\Lambda$.

At this point, it is important to understand the physical meaning of the {\it Fisher quantum tilt} $\alpha$. From the very definition of the quantum Fisher we observe that the quantum tilt $\alpha$ is determined by {\it the non trivial dependence of the quantum phases defining the de Sitter "vacuum", on the energy scale i.e. on $\Lambda= Hk\eta$}. Moreover, we easily observe that this dependence of the quantum phase on $k\eta$ disappears in the super horizon limit $k\eta \sim=0$.

Before going into the computation of the quantum Fisher tilt let us make some brief remarks on the former correspondence. The correspondence links the tilt in a quasi de Sitter frame defined by a quasi de Sitter energy scale with the quantum Fisher tilt defined in {\it pure de Sitter} for the state $|\Lambda\rangle$ defined at the same energy scale. In this sense the family of different de Sitter vacua for different energy scales encodes, in what concerns the tilt, the different quasi de Sitter states we normally design using some inflaton potential. Note that the former correspondence establishes a relation between:

{\it The vacuum de Sitter energy scale $\Lambda$ and the quasi de Sitter scale at which we measure the tilt.}

\subsection{The quantum tilt}

\begin{figure}
\centering
\includegraphics[width=.9\columnwidth]{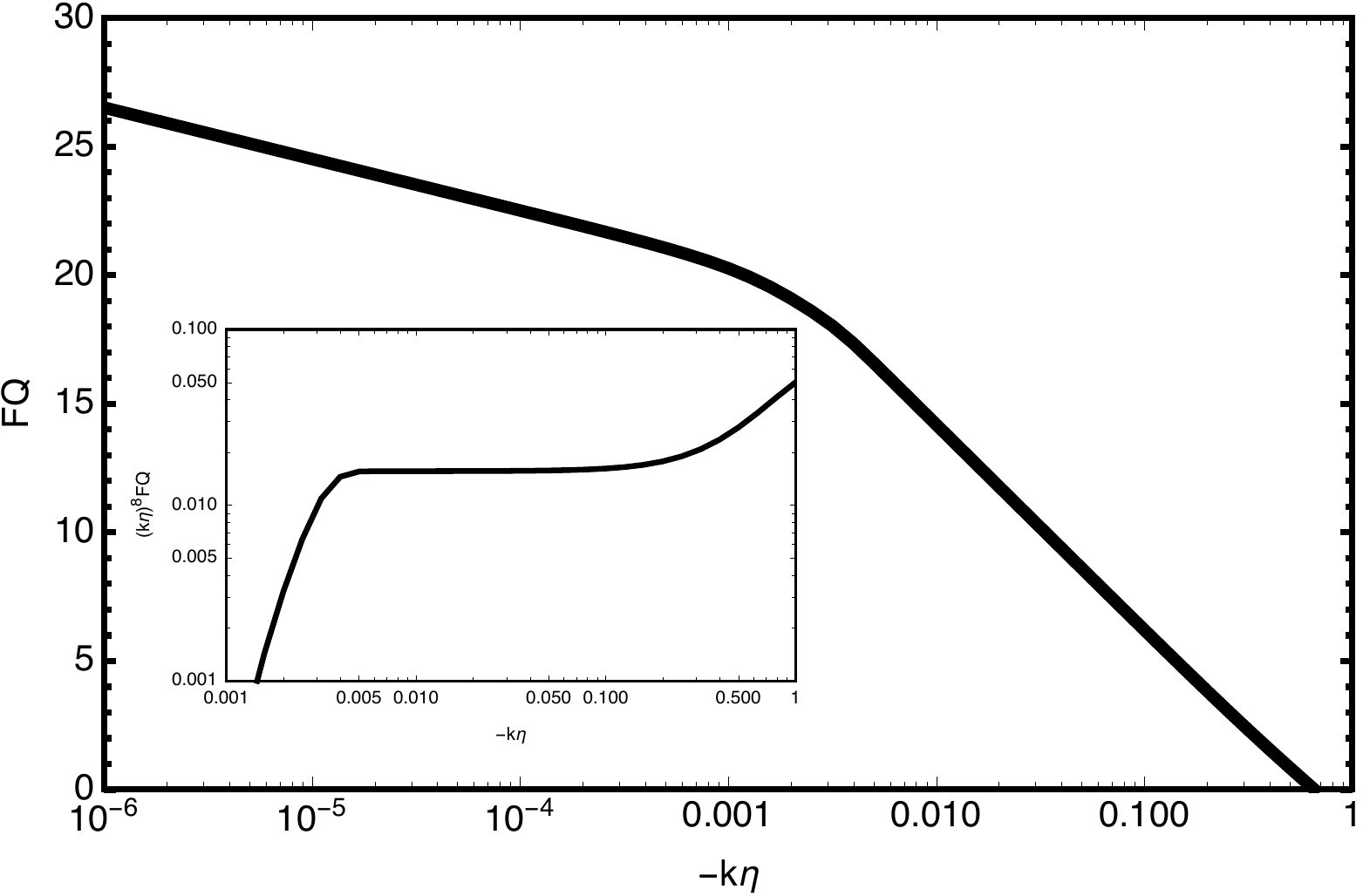} 
\caption{Quantum Fisher ${\cal{F}}_Q$ as a function of $k \eta$. Note the knee at which the power-law behaviour changes slope for $k \eta \sim 0.003$. This corresponds to when the ${\cal{F}}_Q$ becomes classical. The inset shows ${\cal{F}}_Q$ compensated by $(k \eta)^8$ to show the regions where it is nearly scale invariant (constant value). } 
\label{fig:qtilt}
\end{figure}

\begin{figure}
\centering
\includegraphics[width=.9\columnwidth]{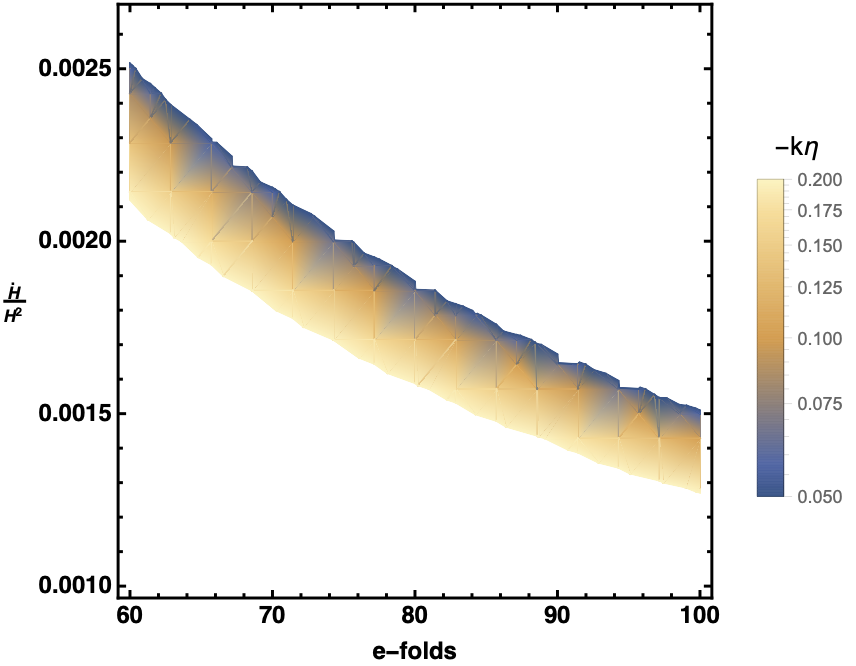} 
\caption{Allowed region for $\frac{\dot H}{H^2}$  and number of e-folds for values of $k \eta$ that are consistent with the Planck18 limits on $1-n_s$.}
\label{fig:banana}
\end{figure}

In this section we will evaluate the tilt of the pure de Sitter quantum Fisher information.
To develop the numerical analysis needed to extract the tilt we redefine 
\begin{equation}
{\cal{F}}_Q \equiv k^2 \eta^4 {F_Q} 
\end{equation}
with $F_Q \equiv {1}{(k\eta)^{8-2\alpha}}$ defined by (\ref{Fisher}). To evaluate the departure from scale invariance we define the ratio
\begin{equation}
{\cal{N}} \equiv ({k \eta})^8 {\cal{F}}_Q
\end{equation}
Fig.~\ref{fig:qtilt} shows $F_Q$ as a function of $k \eta$. Note that it remains a power law in the full range of $k \eta$ but changes slope at around $k \eta  \sim 0.003$. Let us first focus on the region with $k\eta > 0.003$ but smaller than $1$. 
The inset in the figure shows $F_Q$ compensated by $(k \eta)^8$. A scale invariant $F_Q$ would be constant in this case.  $F_Q$ is nearly scale invariant in the range $0.003 < k \eta < 0.5$ (almost close to horizon crossing).
However as can be seen in Fig.~\ref{fig:ns} there is a tilt dependence with scale.
The {\it quantum Fisher spectral index} is defined by this tilt as
\begin{equation}
n_s-1 = 1/2\frac{d\log_{10} ({\cal{N}})}
{d\log_{10} (| k \eta | )}
\end{equation}
Note that this quantum Fisher spectral index defines a {\it red tilt}.

The result is summarized in Fig.~\ref{fig:ns} that shows the resulting tilt as a black line as a function of scale $-k \eta =\frac{\Lambda}{H}$. This {\it quantum tilt} has a scale dependence, thus in order to make contact with experiments we need to fix the scale in the sense of equation (\ref{correspondence}). We are interested in the region with $|k\eta| <1$. For each $k\eta$ the energy scale is given by $\Lambda = k\eta H$ and the ratio relative to horizon exit scale $H$ i.e. $k\eta=1$ is defined as
\begin{equation}
\gamma \equiv \frac{\Lambda}{H}
\end{equation}
Let us now identify the energy scale $\Lambda$ at a given $k\eta$ with the quasi de Sitter value of the Hubble scale after a certain time $\Delta$ counting from the starting moment of inflation. In other words define $\Lambda(\Delta)$ as follows:
\begin{equation}
\Lambda(\Delta) = H(1-\Delta \epsilon H)
\end{equation}
where we use {\it quasi de Sitter slow roll parametrization}. Now we fix the relevant scale $k\eta$ by the correspondence
\begin{equation}
\Lambda(k\eta) = \Lambda(\Delta = n \frac{1}{H})
\end{equation}
for $n$ the number of e foldings before reheating starts. This leads to the phenomenological relation
\begin{equation}
\gamma= (1-n\epsilon)
\end{equation}
that we can interpret as {\it a quasi de Sitter parametrization}. With this prescription we can go to Fig.~\ref{fig:ns} and select the region in parameter space $n,\epsilon$ consistent with the experimental data
\begin{equation}
\alpha(\gamma) \sim 0.05
\end{equation}
The result is shown in Fig.~\ref{fig:banana}. In other words the correspondence defined in \ref{correspondence} becomes:

{\it The quantum tilt at scale $\Lambda$ is in correspondence with the quasi de Sitter tilt for a quasi de Sitter model with $\epsilon$ and $n$ satisfying $\frac{\Lambda}{H} \sim (1-n\epsilon)$.}

From this point of view the possible quasi de Sitter (inflaton potential models) selected by our approach are those that agree (at the level of the tilt) with the  pure quantum, model independent and scale dependent Fisher tilt. As discussed in the introduction, when comparing with the original Mukhanov-Chibisov tilt we find in the region of interest perfect agreement between the tilt dependence on the quasi de Sitter scale $M$ and the energy dependence of the quantum tilt. 

As a final comment we would like to address the attention of the reader to the "crossover" in the behavior of the quantum Fisher function for $k\eta \leq 0.003$. In this region the slope of the quantum Fisher becomes order $4$ which is exactly the behavior of the {\it classical Fisher information}. The potential deep meaning of this crossover deserves further investigation\footnote{A fascinating possibility is that this crossover in the behavior of the quantum Fisher function defines an upper bound on the potential duration of inflation.}

\section{Summary}

The main message of this article is that the primordial tilt for curvature fluctuations that triggers the large scale structure of the Universe can be extracted using as data a primordial de Sitter geometry and the quantum Fisher information measuring the energy scale dependence of the vacuum. At first sight, this is very surprising since it replaces the classical models of decay of the primordial dark energy responsible for inflation --that we normally define using different forms of inflaton potential-- by the intrinsic quantum properties of de Sitter. In the approach we have presented here, these quantum features of de Sitter vacuum are simply characterized by the quantum Fisher information of the  $\Lambda$ vacua. What we find remarkable is that this quantum Fisher information contains the departure from scale invariance we need to account for a realistic spectral index and that the value we find, in a completely model-independent way, agrees with the observed one. More precisely: the pure quantum tilt of Fisher at vacuum scale $\Lambda$ defines the power spectrum tilt at the corresponding quasi de Sitter scale. In a certain sense the former findings make the cosmological tilt a very robust {\it purely quantum and model independent effect} and the quantum Fisher information of de Sitter a good fine grained information of the observed departure from a scale invariant spectrum.

\acknowledgments
 The work of CG was supported by grants SEV-2016-0597, FPA2015-65480-P and PGC2018-095976-B-C21. The work  of RJ is supported by MINECO grant PGC2018-098866-B-I00 FEDER, UE.

\end{document}